\pdfoutput=1
\documentclass{emulateapj}
\newcommand{\kms}{\ensuremath{\mathrm{km\ s}^{-1}}}

\newcommand\htwo{H$_{\rm 2}$}

\newcommand{\Sub}[1]{_\mathrm{#1}}

\renewcommand\d{\Sub{d}}
\renewcommand\H{\Sub{H}}
\newcommand\B{\Sub{B}}

\newcommand\cmq{$\rm cm^{-3}$}
\newcommand\twomic{2.12$\mu$}

\newcounter{ionstage}
\renewcommand{\ion}[2]{\setcounter{ionstage}{#2}%
  \ensuremath{\mathrm{#1\,\scriptstyle\Roman{ionstage}}}}

\newcommand\vtan{V$\rm _{tan}$}
\newcommand\vs{V$\rm _{s}$}
\newcommand\Vr{V$\rm _{r}$}
\newcommand\phis{$\phi \rm _{s}$}

\slugcomment{Accepted for publication in the Astronomical Journal}

\shorttitle{Motions and Spectroscopy within the Ring Nebula}
\shortauthors{O'Dell et al.}

\begin{document}

\title{Tangential Motions and Spectroscopy within NGC 6720, the Ring Nebula\footnotemark[1]} 
\footnotetext[1]{Based on observations with the NASA/ESA Hubble Space Telescope,
obtained at the Space Telescope Science Institute, which is operated by
the Association of Universities for Research in Astronomy, Inc., under
NASA Contract No. NAS 5-26555 and the San Pedro M\'artir Observatory
operated by the Universidad Nacional Aut\'onoma de M\'exico.}

\author{C. R. O'Dell}
\affil{Department of Physics and Astronomy, Vanderbilt University, Box 1807-B, Nashville, TN 37235}
\author{W. J. Henney}
\affil{Centro de Radioastronom\'{\i}a y Astrof\'{\i}sica, UNAM Campus Morelia, Apartado Postal 3-72,
58090 Morelia, Michaoac\'an, M\'exico}
\and
\author{F. Sabbadin}
\affil{INAF-Astronomical Observatory of Padua, Vicolo dell'Osservatorio 5, 35122 Padua, Italy} 
\email{cr.odell@vanderbilt.edu}

\begin{abstract}
We have combined recent Hubble Space Telescope WFPC2 images in the [O~III] 5007 \AA\ and [N~II] 6583 \AA\ lines with similar images made 9.557 years earlier to determine the motion of the Ring Nebula within the plane of the sky. Scaled ratio images argue for homologous expansion, that is, larger velocities scale with increasing distance from the central star. The rather noisy pattern of motion of individual features argues for the same conclusion and that the silhouetted knots move at the same rate as the surrounding gas.  These tangential velocities are combined with information from a recent high resolution radial velocity study to determine a dynamic distance, which is in basic agreement with the distance determined from  the parallax of the central star. We have also obtained very high signal to noise ratio moderate resolution spectra (9.4 \AA) along the major and minor axes of the nebula and from this determined the electron temperatures and density in the multiple ionization zones present. These results confirm the status of the Ring Nebula as one of the older planetary nebulae, with a central star transitioning to the white dwarf cooling curve.

\end{abstract}


\section{Introduction}

The study of planetary nebulae (PNe) has evolved enormously over the last half century, largely through the developing of new observational techniques. However, some fundamental subjects within the field have changed only slightly. Outstanding among these latter subjects are distances and the three dimensional (3-D) structure of individual PNe.  Unfortunately, only a few PNe are close enough for accurate determination of trigonometric parallaxes \citep{har07} and the determination of a 3-D model from high velocity slit spectroscopy has largely been limited to a narrow range of relatively  high surface brightness but well resolved PNe, c.f.~Sabbadin et~al.\@ (2006) and references within. In the case of the closest classical PN (NGC 7293, the Helix Nebula) there is an accurate trigonometric distance and in a series of papers utilizing imaging and spectroscopy a quadrapolar model was derived \citep{omm04,mea08}. Another classical nebula, NGC 6720, the Ring Nebula, was the subject of 3-D modeling \citep{ode07b}  through extensive spectroscopic mapping, but in this case the trigonometric parallax distance of 700$\pm ^{450}_{200}$ pc \citep{har07} is quite uncertain, as are some properties of the central star. Fortunately, if a good kinematic 3-D model of a PN is available, measurements of tangential motions can be used to determine distances. This has been done for a large number of PNe \citep{ter97} using VLA observations of the radio continuum with spatial resolutions of about 1\arcsec, but the necessary supporting kinematic models are usually not available and a defacto spherical model was assumed in deriving most of the  distances. There is also the uncertainty resulting from the radio continuum observations sampling a different portion of the nebula than that studied in the optical emission lines that were used for deriving expansion velocities and the problems associated with the fact that radial velocities measure bulk motion of the gas whereas apparent expansion also includes the effect of the changing position of the ionization front.

O'Dell et al.\@ (2007b) determined the 3-D structure of NGC 6720 from optical line spectroscopy within the assumption of homologous (resembling ballistic)  expansion, a general characteristic of classical PNe known for more than half a century \citep{wil50,wee68} even though such a general expansion characteristic is difficult to theoretically understand \citep{per04,ode07b,sab08}. A recent study \citep{ste08} demonstrates the errors that can arise from an incorrect assumption of homologous expansion, but unfortunately the probable errors in our current study do not allow questioning that assumption in any detail. In the O'Dell et~al.\@ (2007) study it was determined that the nebula was a triaxial ellipsoid seen nearly pole-on, with most of the mass confined to an elliptical ring lying almost in the plane of the sky. The basic correctness of the model was confirmed from application of a very different method of 3-D modeling in a study by Steffen et~al.\@ (2007), which also added modeling of the outer halos of the Ring Nebula. However, the conclusions drawn from the radial velocity study were not as secure as one would hope because radial velocities of a pole-on ellipsoid with most of the material nearly in the plane of the sky are not sensitive ways of mapping where the material is located and of the material's spatial velocity. This is the underlying reason for formulating an observing program that will add information about the tangential velocities. 

In this article we report on a study of the Ring Nebula that employed imaging observations of an order of magnitude higher spatial resolution than the VLA studies of other PNe, but comparable or longer time baselines. Moreover, we utilized  the same optical lines for both spectroscopy and imaging.  The imaging observations and their analysis are described in \S\ 2.1 and \S\ 3. New spectrophotometric observations are described in \S\ 2.2  and their use for determining corrections to the apparent tangential expansion velocities is described in \S\ 4. These combined data, together with the results of our earlier spectroscopic study  \citep{ode07b}, are the basis of the discussion presented in \S\ 5.

\section{Observations}

\subsection{HST WFPC2 Imaging}
We obtained new Hubble Space Telescope (HST) WFPC2  images as parts of programs GO 11231 and GO 11232. The GO 11231 observations were in the F469N (He~II), F547M (continuum), and F673N ([S~II]) filters and executed on 2008 April 23. Triple exposures of 60, 200, and 260 seconds respectively were combined to provide high signal to noise ratio images free from cosmic ray events. The GO 11232 observations were made in the F658N ([N~II]) and F502N ([O~III]) filters on 2008 May 6. Again triple exposures of 260 seconds were made in each filter and similarly combined using IRAF tasks \footnote{IRAF is distributed by the National Optical Astronomy Observatories, which is operated by the Association of Universities for Research in Astronomy, Inc.\ under cooperative agreement with the National Science foundation.}. Identical pointing was employed with the central star of the nebula centered in CCD3 of the WFPC2, duplicating as close as possible the conditions of earlier observations in F502N and F658N in program GO 7632, except that the current two-gyro pointing restrictions of the HST caused us to adopt a nearly reciprocal position angle (P.A.) of 213.2\arcdeg, while the GO 7632 observations were made with P.A.=33.8\arcdeg. The interval between the GO 7632 and GO 11232 observations in the F502N and F658N filters used for astrometry is 9.557 years. The GO 11232 images were aligned with the geomap and geotran IRAF tasks using four stars, including the central star. The alignment of the central star was corrected with the imshift task to an accuracy of better than 0.02\arcsec, leaving the possibility of a poorer alignment accuracy in rotation for objects further out in the nebula, but eliminating any effects of the other field stars not moving in the same manner as the central star.

\subsection{Moderate Resolution Spectroscopy}

The Ring Nebula has been the target of many spectroscopic programs at various spectral resolutions,
the latest and most comprehensive high resolution study being that of O'Dell et~al.\@ (2007b), while recent moderate resolution studies are those of  Guerrero, et~al.\@ (1997), Garnett \&\ Dinerstein (2001), and Liu et~al (2004).  Moderate resolution spectroscopy offers the opportunity to measure the most useful diagnostic emission lines, but previous studies have not combined study of the critical position angles (the major axis points towards P.A.=60\arcdeg\ and the minor axis points towards P.A.=150\arcdeg) and full wavelength coverage. We have closed this information-gap by making new observations with the Boller and Chivens Spectrograph mounted on the 2.12 m telescope at the San Pedro Martir Observatory.  

It was expected that these observations would also provide ground based telescope reference data for calibrating the F469N and F673N filters of the HST's WFPC2, which meant that we used an unusually wide slit (in order to diminish the effects of comparing high and low spatial resolution data). We employed a 300 $\mu$m wide entrance slit (3.93\arcsec). The scale of the Site CCD was 1.05\arcsec\ pixel along the slit and 3.05 \AA/pixel along the dispersion direction.  The full width at half maximum signal (FWHM) for the final combined spectra along the direction of the dispersion was 9.4 \AA.

Observations were made at P.A.=150\arcdeg\ on the night of 2008 July 25 and at P.A.=60\arcdeg\ on the night of 2008 July 26 under photometric conditions. The flux standard BD+25~3941 in the IRAF database was used for both nights.  Various exposure times were used in order to avoid the problems of the bright lines saturating on the longer exposures. For P.A.=150\arcdeg\ we obtained seven exposures of 180 seconds, five of 600 seconds, and four of 1200 seconds for a total exposure time of 9060 seconds. For P.A.=60\arcdeg\ we obtained three exposures of 180 seconds, three of 600 seconds and two of 1200 seconds for a total exposure time of 4740 seconds. The seeing FWHM was about 2.3\arcsec\ on both nights. Twilight sky exposures were made on the evening of the P.A.=150\arcdeg\ observations and were used for characterizing the small flat field corrections.  

The data were reduced using IRAF tasks, which included flat field correction, distortion correction, wavelength calibration, and rendering into cgs system surface brightness units (ergs s$^{-1}$ cm$^{-2}$ \AA $^{-1}$ steradian$^{-1}$). The spectra were combined to form a single spectrum for each position angle, with only the shorter exposure time observations used for the brightest emission lines. Digital versions of these spectra are available in the electronic form of this paper.

\begin{figure}
\epsscale{0.8}
\plotone{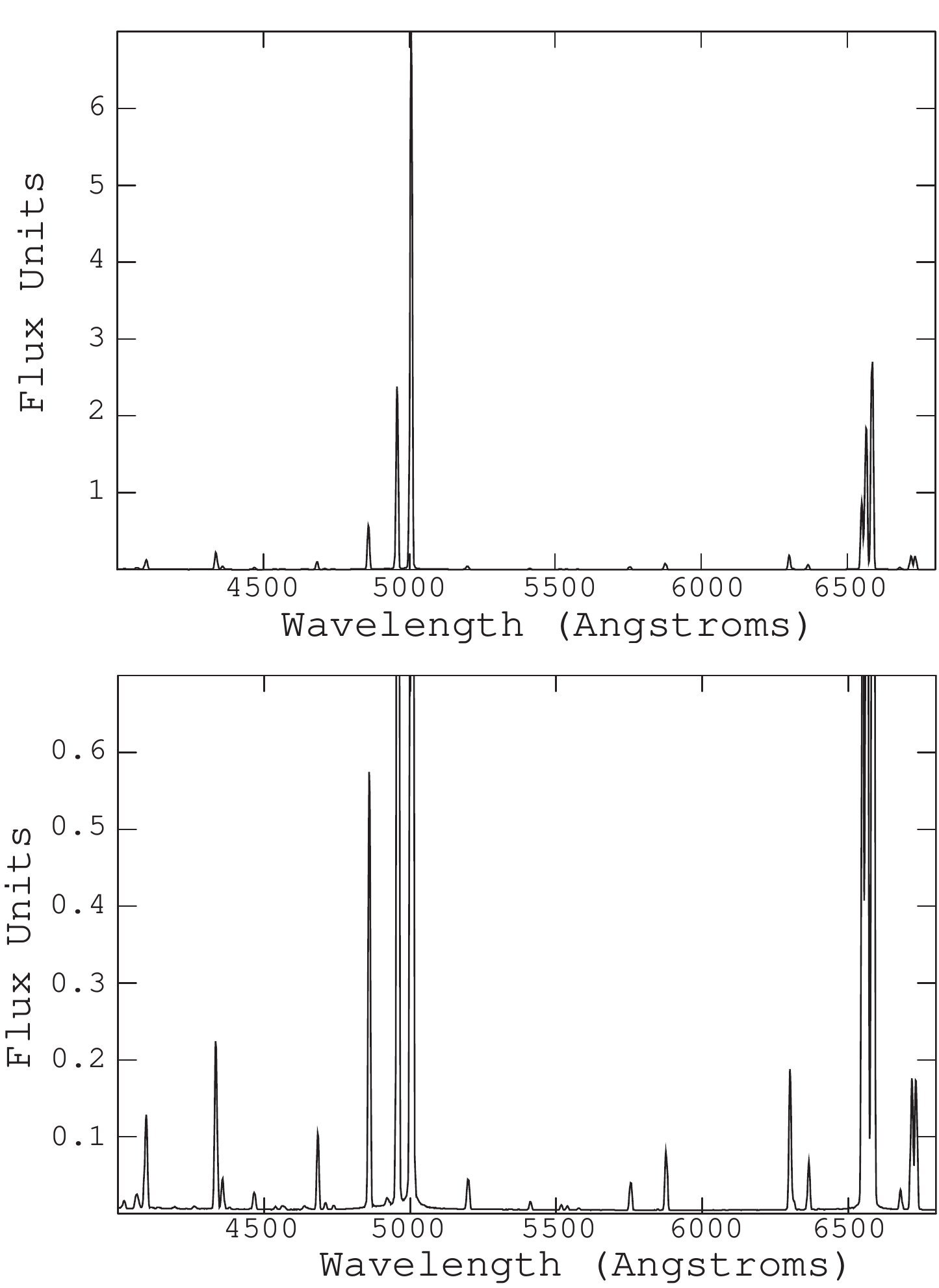}
\caption{A representation of the spectra is shown. The particular sample is PA150top. These are observed (not dereddened) values in relative flux units, the maximum value of the upper panel is ten times that of the lower panel.}
\label{ratios}
\end{figure}

\begin{deluxetable*}{lllllllll}
\tabletypesize{\scriptsize}
\tablecaption{Extinction Corrected Line Flux Ratios}
\tablewidth{0pt}
\tablehead{
\colhead{Wavelength (\AA)} & \colhead{Ion} & \colhead{f$_{\lambda}$} & \colhead{PA150top} & \colhead{PA150bot} & \colhead{PA150mid} & \colhead{PA60top} & \colhead{PA60bot} & \colhead{PA60mid}}
\startdata
4026 & HeI,HeII & 0.29     &   2.33 &   2.21 &   1.59 &   2.25 &   2.86 &   2.66\\
4068 & [S~II]* & 0.28     &   4.42 &   5.58 &   3.96 &   4.71 &   5.58 &   2.40\\
4102 & H$\delta$ & 0.27     &  26.2  &  26.1  &  23.8  &  25.4  &  27.5  &  23.1 \\
4267 & O~II & 0.19     &   0.73 &   0.71 &   1.79 &   0.74 &   0.82 &   1.39\\
4340 & H$\gamma$ & 0.16     &  43.0  &  43.2  &  44.2  &  43.1  &  44.9  &  43.9 \\
4363  & [O~III] & 0.16     &   7.93 &   7.50 &   9.64 &   6.93 &   6.32 &   9.59\\
4471 & HeI & 0.12     &   4.18 &   4.29 &   2.86 &   4.62 &   4.73 &   2.82\\
4540 & HeII & 0.11     &   0.63 &   0.56 &   1.69 &   0.35 &   0.42 &   1.65\\
4686 & HeII & 0.04     &  16.5  &  15.2  &  49.3  &  11.6  &  11.6  &  45.5 \\
4711& [Ar~IV]*& 0.02     &   1.62 &   1.31 &   4.52 &   1.04 &   1.04 &   4.15\\
4739 & [Ar~IV]* & 0.02     &   0.95 &   0.82 &   2.98 &   0.52 &   0.48 &   2.79\\
4861 & H$\beta$ & 0.00     & 100.0  & 100.0  & 100.0  & 100.0  & 100.0  & 100.0 \\
4922 & HeI &-0.02     &   2.58 &   1.62 &   0.84 &   1.35 &   1.32 &   1.00\\
4959 & [O~III] &-0.02     & 405.5  & 260.6  &  361.4 & 376.7  & 380.2  & 397.2 \\
5199 & [N~I] &-0.10     &   6.90 &   7.50 &    0.71&   8.81 &   8.87 &   0.80\\
5411 & HeII &-0.14     &   1.85 &   1.60 &    5.08&   1.21 &   1.22 &   4.55\\
5517 & [Cl~III] &-0.17     &   1.03 &   1.07 &    0.62&   0.87 &   0.91 &   0.57\\
5537 & [Cl~III] &-0.17     &   0.82 &   0.86 &    0.47&   0.70 &   0.74 &   0.43\\
5577 & [O~I] &-0.19     &   0.35 &   0.35 &    -   &   1.21 &   0.13 &   0.57\\
5593 & O~III &-0.19     &   -    &     -  &   0.022 &    -   &   -    &   0.021\\
5755 & [N~II] &-0.21     &   5.86 &   6.30 &   0.90&   6.35 &   6.30 &   1.09\\
5876 & HeI &-0.23     &  12.4  &  12.6  &   8.07 &  13.1  &   13.1 &  7.80\\
6300 & [O~I] &-0.31     &  29.58 &  31.4  &  2.00 &   36.8  &   37.2 &  2.70\\
6363 & [O~I] &-0.31     &   9.86 &  10.6  &   0.63 &  12.3  &   12.4 &  0.86\\
6548 & [N~II] &-0.35     & 136.8  & 151.0  &  21.1 &  156.0  &  156.0 & 23.4\\
6563 & H$\alpha$ &-0.35     & 292.4  & 294.4  &  288.4 &  283.8 &  284.4 & 269.2\\
6583 & [N~II] &-0.35     & 421.7  & 458.1  &   64.3 &  469.9 &  469.9 &  69.0\\
6678 & HeI &-0.37     &   4.00 &   4.13 &   2.96 &   5.06 &   5.04 &  3.29\\
6717 & [S~II] &-0.38     & 26.3   &  29.7  &   5.48 &  24.3  &  24.4  &  5.12\\
6731 & [S~II] &-0.38     & 29.1   &  32.9  &   6.07 &  26.9  &  27.0  &  5.66\\
\enddata
\tablecomments{~The ratios are normalized to H$\beta$ =100. 4068 is likely to be contaminated by O~III lines in regions of very high ionization, 4711 can be contaminated by HeI lines in regions of low ionization and [Ne~IV] lines in regions of high ionization, 4739 is subject to contamination by [Fe~III] and [Ne~IV] lines.}
\end{deluxetable*}

\begin{deluxetable*}{lcllccll}
\tabletypesize{\scriptsize}
\tablecaption{Electron Temperatures* and Densities*}
\tablewidth{0pt}
\tablehead{
\colhead{Sample} & \colhead{pixel offset} & \colhead{T$_{e}$([O~III])} & \colhead{T$_{e}$([N~II])} & \colhead{T$_{e}$([S~II])} &\colhead{T$_{e}$([O~I])} & \colhead{N$_{e}$([Cl~III])} & \colhead{N$_{e}$([S~II])}}
\startdata
PA150top & 11 -- 31 &1.01 & 1.01 & 0.81 & 0.81 & 600 & 860\\
PA150bot & -14 -- -36 &1.15 & 1.00 & 0.88 & 0.80 & 850 & 860\\
PA150mid & 3 -- 10, -3 -- -13 &1.12 & 1.01 & - & - & 460 & 860\\
PA60top & 19 --- 46 & 0.99 & 0.99 & 0.86 & 1.19 & 850 & 860\\
PA60bot & -20 -- -46 & 0.96 & 0.99 & 1.02 & 0.61 & 370 & 860\\
PA60mid &  3 -- 18, -3 -- -19 & 1.08 &1.06 & - &  - & 410 & 860\\
\enddata
\tablecomments{~Electron temperatures (T$_{e}$) are given in units of 10$^{4}$ K and densities (N$_{e}$) in units of \cmq.}
\end{deluxetable*}

Night sky lines were present as were the high ionization lines arising from the outer halo of the Ring Nebula. The regions beyond the bright ring were used to determine how much should be subtracted from the ring spectra. A final pair of background subtracted spectra of 107 pixels length centered on the bright ring was created. Both of these spectra were sampled in four regions called top (25 pixels) , bottom (bot, 25 pixels), and middle (mid, 13 pixels) with the results presented in Table 1. An illustration of the quality of the spectra is shown in Figure 1. The range of pixels measured from the central star, with positive numbers away from the direction of the P.A., are given in Table 2. The pairs of middle samples were adopted because of the much lower surface brightness near the central star. The sample region spectra were measured with IRAF task ``splot'', with deconvolution of line blends where necessary. Not all features were measured, in particular those blended lines that would not be useful in a quantitative analysis.  The interstellar extinction correction was determined from the F(H$\alpha$)/F(H$\beta$) and F(H$\delta$)/F(H$\beta$) ratios using the general extinction curve presented in Osterbrock \&\ Ferland (2006) and their theoretical line ratios of 2.87 and 0.256 respectively. The weighted average extinction was 
c$\rm _{H\beta}$=0.13$\pm$0.04 (in close agreement with the value of 0.14 derived by Guerrero, et~al.\@ (1997) but about half that of Barker (1987)) and the same extinction correction was applied to each sample. The extinction curve utilized and the results of this analysis of the samples are presented in Table 1.  

We also derived c$\rm _{H\beta}$ pixel by pixel along both slits. This indicated that there is a sharp rise just outside of the peak in H$\beta$ brightness in the directions P.A.=150\arcdeg, 240\arcdeg, and 330\arcdeg, but not towards 60\arcdeg. These rises are like those shown in Garnett \&\ Dinerstein (2001) towards P.A.=271\arcdeg, although they questioned the reality of the rise because it might be due to misalignment of their spectra, which is not an issue with our observations. Since the peaks in brightness occur just inside the projection of the ionization front and the front is curved, the rise in extinction just outside the peaks is probably due to dust in dense neutral gas projected onto ionized material inside the ionization front. 

Our spectra are comparable in resolution to the 7 \AA\ resolution spectra of Guerrero, et~al.\@ (1997) at similar P.A. values and with a slit width of 1.2\arcsec, however, their exposure time along the major axis was 1800 sec (ours was 4740 sec) and the sum of the exposure times of their spectra nearly along the minor axis was 4500 sec (ours was 9060 sec), although it should be noted that they covered a wider wavelength range (3727 \AA\ through 7330 \AA).  Our resolution was poorer than the 2.1-2.3 \AA\ reported by Garnett \&\ Dinerstein (2001), who had a total exposure time of 3600 sec at P.A.=91\arcdeg\ with a 2.5\arcsec\ wide slit, but covered a wider wavelength range than their 4150 \AA\ through 5000 \AA\ sample.  The study of Liu et~al.\@ (2004) was made with a P.A.=90\arcdeg\ slit at a resolution of 1.5 \AA\ in their blue samples and 3.5 \AA\ in the red, both with a 1" slit and a maximum total exposure time in their reddest sample of 2560 sec. They reduced the long-slit spectrum to a single spectrum, which we show in \S\ 4.1 to blend regions of quite different conditions. Their total wavelength coverage was 3600 \AA\ through 8010 \AA. Because of the wider entrance slit and longer exposure times, our data are of much higher spectrophotometric quality than these earlier studies, within the limitations of the spectral resolution. A measure of the large signal in this program is that the 4922  \AA\ line of HeI in the south east peak of the main ring gave an individual 1200 sec exposure signal of 850 counts in one pixel along the slit, even though this line is only about 1\%\ the flux of the H$\beta$ line and this observation represents only 13 \%\ of the entire signal for the 4922 \AA\ line.

Since the central star has a surprisingly large spread in published V magnitude brightness (15.0--16.2)  \citep{ode07b}, we carefully measured this star on our spectra at 5550 \AA (near the center of the V bandpass) and determined that the correct value is V=15.65$\pm$0.05, in good agreement with the value of 15.75 determined by Harris, et~al.\@ (2007). The small difference in our new results and the value adopted in our earlier study of the nebula \citep{ode07b} mean that the stellar properties adopted in that paper do not need to be changed.

\section{Derivation of Tangential Velocities}

In an earlier study of the Ring Nebula \citep{ode02} pairs of HST WFPC2 images in the F658N filter were compared by magnifying the first epoch images until the difference of the first and second images was a minimum. The time interval between the images employed was 1.944  years and the magnification necessary was 1.0013$\pm$0.0002. This corresponds an annual expansion rate (motion in mas yr$^{-1}$ divided by the distance $\phi$ from the central star in arcseconds) of 0.7$\pm$0.1 mas yr$^{-1}$/\arcsec.  However, a later paper \citep{ode07b} argued that these results were possibly incorrect since the central intersection of the WFPC2  centered on the central star, which meant that the image of the nebula fell into four different CCD's and the known drift of the CCD's with respect to one another \citep{and03} meant that an internally self consistent but flawed answer could be obtained. It is for this reason that we formulated the GO 11232 observing program, where the images fell into a single CCD, whose results we present here. Our new data were processed in two ways, first by magnification and comparison, as done before, and by measurement of individual features.

\subsection{Comparison of Scaled Images}
\begin{figure*}
\includegraphics[width=\linewidth]{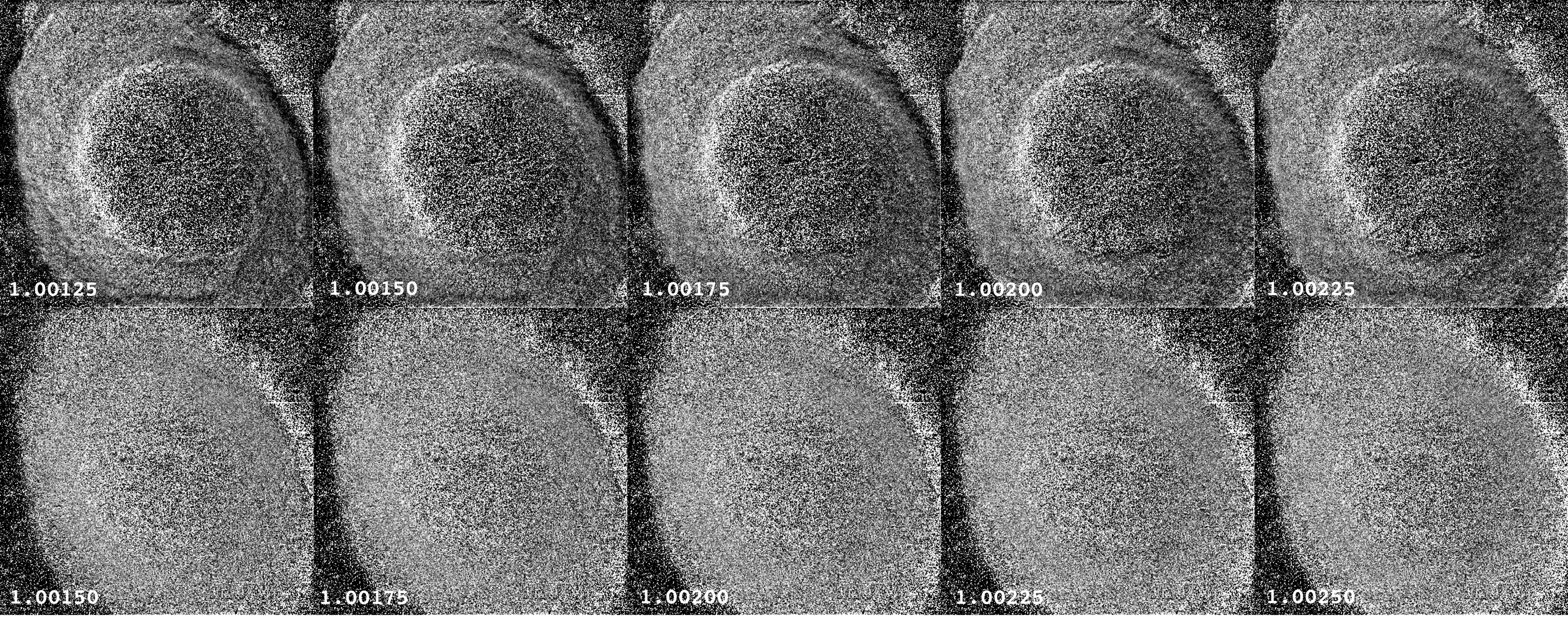}
\caption{These 70.0\arcsec x68.6\arcsec\ ratio images of the Ring Nebula have the vertical axis pointed towards P.A.=33.8\arcdeg. They were generated by scaling the aligned first epoch images by the factors shown, then divided by the second epoch images. The top row is of the F658N images and the lower row of the F502N images.  The display range of all images are 0.95 to 1.05.}
\label{ratios}
\end{figure*}

Our method of comparison was to magnify the first epoch images, align these with the second epoch image in the same filter, normalize them so that the total signal was the same, then take their ratio. We noted that the ratio images began as being highly structured, indicating a mismatch, then became nearly grey in a small scaling range, then again became structured at larger scaling factors. Samples of the ratio images used in the analysis are shown in Figure~2. This rather qualitative method gave a best scaling factor of 1.0016$\pm$0.0005 for the F658N images and 1.0019$\pm$0.001 for the F502N images, yielding annual expansion rates  over the 9.557 year interval of 0.17$\pm$0.05 mas yr$^{-1}$/\arcsec\ and 0.20$\pm$0.05 mas yr$^{-1}$/\arcsec\ respectively, a weighted average (F502N being given one half weight because of less structure in that filter) of 0.18$\pm$0.05 mas yr$^{-1}$/\arcsec,  only a fraction of the annual expansion rates previously derived from segmented images and arguing that those rates are in error because of the use of images divided onto four CCD's. The probable errors of our present study are very qualitative as they depend strictly on judgement about when the fit is best.  Non-homologous expansion would have manifested itself by different parts of the ratio image becoming featureless at different scale factors. Qualitatively, this did not seem to be the case except in the bright northern quadrant, where some non-homogeneity could be seen in the inner and outer parts of the brightest section (which was homologous). 

\subsection{Measurement of Individual Features}
\begin{figure*}
\epsscale{1.0}
\plotone{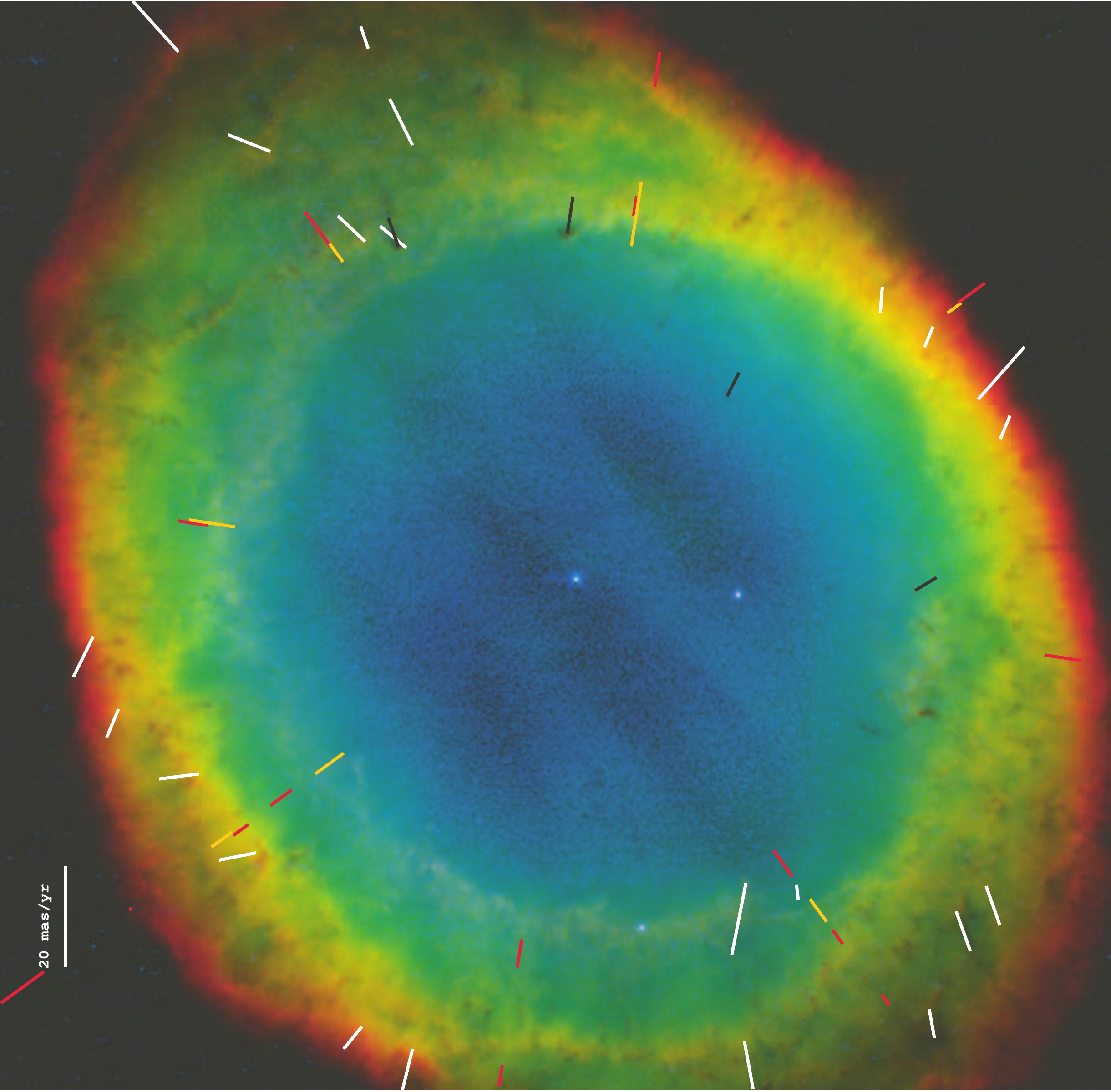}
\caption{The same field of view as Figure 2 is shown in color with lower ionization stage emission portrayed as longer wavelengths: Blue-HeII 4686 \AA, Green-[O~III] 5007 \AA, and Red-[N~II] 6583 \AA. Measured motions within the nebula are shown by lines (white for [N~II] bright knots, black for [O~III] knots seen in silhouette, red for nebular features in [N~II], and yellow for nebular features in [O~III]) with the length of the line indicating the degree of motion and the inner point the position of the feature measured. 
\label{motions}}
\end{figure*}

\begin{deluxetable*}{ccccc}
\tabletypesize{\scriptsize}
\tablecaption{Measured Expansion Motions of Knots in the F658N Filter}
\tablewidth{0pt}
\tablehead{
\colhead{P.A.$\rm _{Object}$ (\arcdeg)} & \colhead{$\phi\rm _{Object}$ (\arcsec)} & \colhead{Motion (mas yr$^{-1}$)}
& \colhead{P.A.$\rm _{Motion}$ (\arcdeg)} & \colhead{Motion/$\phi$ (mas yr$^{-1}$/\arcsec)}}
\startdata
54.7  &  29.3 & 10.20   &    60.1     &   0.348\\
55.4  &  35.9 &  4.65   &    52.4     &   0.130\\
61.2  &  23.5 &  6.69   &    84.1     &   0.285\\
66.0  &  25.1 &  7.37   &    80.6     &   0.294\\
69.4  &  33.2 &  8.98   &   102.2     &   0.270\\
70.9  &  41.7 & 13.86   &    75.8     &   0.332\\
130.6 &  30.8 &  7.90   &     127.7   &   0.256\\
139.6 &  30.1 &  6.12   &     131.1   &   0.203\\
164.2 &  26.6 &  7.42   &     135.0   &   0.279\\
177.1 &  26.9 &  8.00   &     156.6   &   0.297\\
189.1 &  31.3 &  5.70   &     175.4   &   0.182\\
233.6 &  31.0 &  9.68   &    228.1    &   0.312\\
248.7 &  21.9 &  1.46   &    202.7    &   0.067\\
253.0 &  35.0 &  5.78   &    224.1    &   0.165\\
262.5 &  31.8 &  8.38   &    241.9    &   0.429\\
266.8 &  32.2 &  8.25   &    243.1    &   0.256\\
322.1 &  28.1 &  5.00   &      1.4    &   0.178\\
327.9 &  27.7 &  2.08   &    352.5    &   0.075\\
337.5 &  26.3 &  4.42   &    357.5    &   0.168\\
345.2 &  25.4 &  5.13   &     29.6    &   0.202\\
\enddata
\end{deluxetable*}

\begin{deluxetable*}{ccccc}
\tabletypesize{\scriptsize}
\tablecaption{Measured Expansion Motions of Knots in the F502N Filter}
\tablewidth{0pt}
\tablehead{
\colhead{P.A.$\rm _{Object}$ (\arcdeg)} & \colhead{$\phi \rm _{Object}$ (\arcsec)} & \colhead{Motion (mas yr$^{-1}$)} & \colhead{P.A.$\rm _{Motion}$ (\arcdeg)} & \colhead{Motion/$\phi$ (mas yr$^{-1}$/\arcsec)}}
\startdata
3.5   & 14.9 &  7.34   &    6.8 & 0.493\\
35.5  & 21.8 &  7.41   &   25.1 & 0.340\\
62.2  & 23.8 &  5.98   &   53.3 & 0.251\\
301.8 & 21.3 &  5.01   &   334.9 & 0.235\\
\enddata
\end{deluxetable*}

The more quantitative method employed compared the position of isolated individual features (bright in the case of F658N and dark in the case of F502N) on aligned, normalized images. The method is one of shifting small sections of an image and identifying when the differences are least, a method previously developed \citep{har01} for measurement of Herbig-Haro features. The results are shown in Figure~3 and are presented in Table 3 and Table 4. In the same manner as the comparison of images method, we can express our results as the motion divided by the distance to the central star, which would be a constant for homologous expansion. The errors in the position for individual features  expressed in relative motion were typically 0.02 mas yr$^{-1}$/\arcsec. For F658N this method gave for 20 objects a global average  for the expansion rate of 0.23$\pm$0.08 mas yr$^{-1}$/\arcsec\  and for four F502N  objects 0.33$\pm$0.12 mas yr$^{-1}$/\arcsec, for a weighted average of 0.24$\pm$0.11 mas yr$^{-1}$/\arcsec. These values are slightly larger than the results from simply comparing the scaled images; however, these are to be preferred because of their more quantitative derivation. In this case individual features are measured and an estimate of their accuracy is determined from the scatter of the results. In the scaling method described in \S\ 3.1 there is only the qualitative judgement of when the scaling is correct, with the uncertainty estimated by when the scaling would become unreasonable. There were no statistically significant differences in the expansion properties in the four quadrants sampled.

We determined by comparing the P.A. of the measured motions with the P.A. from the central star that possible slight misalignment of the images in rotation did not strongly affect our results. We found that on the average the motion vectors were 5\arcdeg\ smaller. This could be the results of a misalignment of the first and second epoch images of only 0.0116\arcdeg\ and this would affect the derived  expansion motions by only about 0.05\%.

\begin{deluxetable*}{ccccc}
\tabletypesize{\scriptsize}
\tablecaption{Measured Expansion Motions of Nebular Features}
\tablewidth{0pt}
\tablehead{
\colhead{P.A.$\rm _{Feature}$ (\arcdeg)} & \colhead{Filter} & \colhead{$\phi \rm _{Feature}$ (\arcsec)}  & \colhead{Motion (mas yr$^{-1}$)} &  \colhead{Motion/$\phi$ (mas yr$^{-1}$/\arcsec)}}
\startdata
25 & F658N & 23.0     &        3.96      &      0.17\\
  -   &     -        & 31.8     &       6.98       &      0.22\\
 70 &   -         &  26.3     &       8.96   &        0.34\\
 115 & - & 23.5 & 6.04 & 0.26\\
 160 & - & 41.7 & 10.53 & 0.25\\
 -       & - & 33.5 & 0.63 & 0.02\\
 -       & - & 25.6 & 3.65 & 0.14\\
 -    & - & 22.4 & 5.21 & 0.23\\
 205 & - & 30.9 & 4.17 & 0.13\\
  -    & - & 22.9 & 5.63 & 0.25\\
  250 & - & 20.9 & 6.46 & 0.31\\
  -     & -  & 26.7 & 3.44 & 0.13\\
  -  &  -  & 31.6 & 2.71 & 0.08\\
  295 & - & 29.6 & 7.40 & 0.25\\
  340 & - & 29.7 & 6.57 & 0.22\\
  25 & F502N & 21.5 & 12.82 & 0.60\\
  70 &  - & 24.9 & 4.48 & 0.18\\
  115 & - & 22.3 & 9.17 & 0.41\\
  160 & - & 26.8 & 5.00 & 0.19\\
   - & - & 18.2 & 6.98 & 0.38\\
   250 & - & 24.9 & 5.52 & 0.22\\
   340 & - & 28.3 & 3.44 & 0.12\\
\enddata
\end{deluxetable*}

We also measured features within the nebula employing a different  approach. The method was to create samples 4.0\arcsec\ wide along P.A. values selected to lie along the major and minor axes of the nebula and intermediate angles displaced by 45\arcdeg.  Averages were taken perpendicular to the P.A. direction and profiles were created. Well defined peaks not associated with knots were then measured in the first and second epoch images and in both filters. Fewer [O~III] features were measured because of the more amorphous appearance of the nebula in the F502N filter. The results are shown in Figure 3 and Table 5.  The expansion rate was 0.30$\pm$0.17 mas yr$^{-1}$/\arcsec\ for F502N, and 0.20$\pm$0.09 mas yr$^{-1}$/\arcsec\ for F658N. Because the F502N features are broader and more difficult to measure, they were assigned half the weight of the F658N features in determining the average expansion rate for the nebular features of 0.22$\pm$0.11 mas yr$^{-1}$/\arcsec.

\section{Analysis of the Spectroscopic Data}

The ionization stratification of the Ring Nebula in the plane of the sky has been well defined by a number of studies including both imaging in various ions \citep{lam94,ode07b} and long-slit spectroscopy like our own \citep{gue97,gar01}. These studies indicate that in the main equator of material, which lies almost in the plane of the sky, that the nebula is optically thick to Lyman continuum radiation (E $\geq$ 13.6 eV). Closer to the central star the radiation field is determined by the optical depth in helium, either singly ionized (E $\geq$ 24.6 eV) or doubly ionized (E $\geq$ 54.4 eV).  As summarized earlier \citep{ode98}, these various zones of ionization have convenient optical emission-line tracers. [O~I] and [S~II] come from the outermost region where hydrogen is transitioning from being highly ionized to being neutral (which we'll call the main ionization front or MIF). [N~II] comes from the next closest region to the central star, where H$^{+}$ and He$\rm ^{o}$ coexist (which we'll call the He$\rm ^{o}$ zone). The next zone is populated by H$^{+}$ and He$^{+}$ (which we'll call the He$^{+}$ zone) and is traced by [O~III] emission.  The innermost zone finds
H$^{+}$ and He$^{++}$ and is traced by the HeII recombination lines like 4686 \AA\ (this region will be called the He$^{++}$ zone). Our spectra cover a sufficiently wide range of line intensities and ionic species, that we can independently determine densities and temperatures for most of the regions. We 
have done this using the IRAF/STDAS/Nebula task ``temden'', which calculates the temperature (for an assumed density) or density (for an assumed temperature) from extinction corrected line ratios for a wide range of ions. 

 It must be kept in mind that in the best 3-D model for the Ring Nebula \citep{ode07b} that the equatorial high density region is elliptical, with its plane lying about 7\arcdeg\ out of the plane of the sky, c. f. \S\ 5.2 and O'Dell, et~al.\@ (2007), and there is a lower density polar region of about 1.5 larger in size that extends perpendicular to the equatorial region \citep{ode07b}. This means that samples in taken the bright ring (our PA150top, PA150bot, PA60top, and PA60bot samples) are measures of what is happening in the equatorial region, while the samples from close to the central star (our PA150mid and PA60mid samples) are seeing radiation from over a wide range of ionization states, with any low ionization emission lines coming from material physically distant from the central star.

\subsection{Determination of the  Physical Conditions}
The electron densities can in principle be determined from doublet line ratios with little dependence on the assumed electron temperature. The [S~II] 6717 and 6731 \AA\ doublet ratio gives densities at the MIF, while the [Cl~III] 5517 and 5537 \AA\ doublet largely arises in the He$^{+}$ zone, with the results shown in Table 2.  Garnet and Dinerstein (2001) used their higher resolution spectra to determine densities in the He$^{+}$ zone from the [Ar~IV] 4711 and 4739 \AA\ ratio, after correcting the 4711 \AA\ for the unresolved HeI 4713 \AA\ line by scaling from the HeI 4471 \AA\ line, finding densities of about 700 \cmq, which are in basic agreement with our [Cl~III] results for this same region.

The electron temperatures can be obtained from ratios of lines having very different excitation levels.
The most widely used is the F (4363)/[F(5007)+F(4959)] of [O~III], which arises in the He$^{+}$ zone. 
Since the 5007 \AA\ line was saturated in the bright ring portion of even our shortest exposure time spectra, we determined the F(5007)/F(4959) from the unsaturated central section of the spectra and found this to be 2.94. The value predicted using the transition probabilities presented in Osterbrock and Ferland (2006) is 2.9 and the observed value could be used to refine the determination of the transition probabilities. The results are also given in Table 2 and are in good agreement with those of earlier studies \citep{bar87,gue97,gar01,liu04}.  The [N~II] F(5755)/[F(6548)+F(6583)] gives temperatures in the He$\rm ^{o}$ zone which are similar to those in the He$^{+}$ zone, which is not surprising since most of the heating arises from Lyman continuum  photons and that radiation field has suffered little modification in these regions.  In even finer sampling of our spectra we see the same rise in the [O~III]  and [N~II] temperatures towards the central star that was found by Garnett \&\ Dinerstein (2001) in [O~III], except that the local peak in temperature at about 10\arcsec\ on both sides of the central star is better defined. This position corresponds to the inner boundary of a peak region of HeII 4686 \AA\ emission, which suggests that there is a central lower density but higher temperature region along the polar axis of the nebula. Temperatures are determined in the MIF by the [F(4068)+F(4076)]/[F(6731)+F(6731)] ratio of [S~II] and the F(5577)/[F(6300)+F(6363)] lines of [O~I].  The intrinsically weak 5577 \AA\ line is absent in our middle region samples, thus precluding derivation of a temperature there and the F(4068)+F(4076) lines are contaminated by line of sight higher ionization lines (in particular O~II recombination lines). 

\subsection{The Ionization Model for the Ring Nebula}
In our earlier paper \citep{ode07b} that produced a 3-D model of the nebula, we presented the results of calculations with the Cloudy \citep{fer98} code using the appropriate stellar and nebular conditions. Excellent agreement was found between the observed and calculated properties for both the axes lying nearly in the plane of the sky. Less good agreement was found for the more poorly defined structure nearly along the line of sight, where the predicted nebular density was too small. This discrepancy is exacerbated by our spectroscopic determination that the densities in the outer parts of these polar regions (as determined from the [S~II] doublet in the middle samples) are no lower than in the outer parts of the planar region. This leaves a quandary about the structure in the polar region, as both the arguments of O'Dell et~al.\@ (2007) and Steffen et~al.\@ (2007) for the ionization of the outer halo call for this region being optically thin.

\section{Discussion}
It is possible to combine the results of our earlier study of the radial velocities \citep{ode07b} together with our new results from spectroscopy and expansion in the plane of the sky to improve our knowledge of the properties of the Ring Nebula. In this section we present the velocity dependence upon the distance from the central star (\S\ 5.1), derive the tilt of the nebula with respect to the plane of the sky (\S\ 5.2), derive the nebula's dynamic distance (\S\ 5.3), comment on corrections that may be necessary due to the relative motion of the MIF with respect to the nebular gas (\S\ 5.4), demonstrate that knots appear to be moving with the expanding nebular gas (\S\ 5.5), then comment on how these results alter the previous picture of the age of the nebula and stage of evolution of the central star (\S\ 5.6).

\subsection{The Radial Velocity Expansion Relation for the Ring Nebula}

Arguments for the existence of homologous expansion in the Ring Nebula appears in both the radial velocity data \citep{ode07b} and in our tangential velocity data. As noted earlier, this type of motion defies explanation by hydrodynamic models, but it cannot be ignored because it has been well established for radial velocities for about half a century \citep{wil50,wee68}. 

In the case of the present study of tangential velocities, the average of the results from measuring motions of the knots and the nebula give a tangential expansion relation of \vtan =0.23$\pm$0.10 x $\phi$ mas yr$^{-1}$, where $\phi$ is the distance from the central star in the plane of the sky in seconds of arc. The relation of this result to the radial velocity expansion relation is determined by the distance to the nebula (D).

By comparing the ratio of radial velocities with respect to the systemic velocity  (\Vr) at various angular distances from the central star, O'Dell et al.\@ (2007b)  determined that \Vr /$\phi$=0.98 \kms/arcsec, without giving an uncertainty. The relation of  \Vr\ to the spatial expansion velocity (\vs) will be determined by the geometry of the spheroid best describing the nebula. O'Dell et al.\@ (2007b) argue that the structure perpendicular to the main ring of the nebula is a spheroid with a vertical length (H) 1.5 times the major axis of the elliptical main ring.  In this case the spatial expansion law would be \vs =0.65x\phis\ \kms, where \phis\ is the angle corresponding to the distance between a point in the nebula and the central star.

\subsection{The Tilt of the Main Body of the Ring Nebula with respect to the Sky}

We can use our new results to more accurately determine the tilt of the plane of the elliptical main ring of the nebula with respect to the plane of the sky ($\theta$). One knows from the non-tilted form of the velocity ellipses of the high resolution spectra made nearly across the minor axis (P.A.=160\arcdeg) that this axis lies in the plane of the sky, whereas those nearly along the major axis (P.A.=70\arcdeg) are tilted with a difference of velocity of 6$\pm$2 \kms, with the east-northeast end approaching the observer.  Assuming that the full major axis is 67\arcsec\ and using the expansion rate of 0.23$\pm$0.10 mas yr$^{-1}$/\arcsec\  means that the tangential velocity is 25.6 \kms. The tangent of the tilt angle will be the ratio of the radial velocity difference (3 \kms) divided by the tangential velocity (25.6 \kms). This yields a tilt angle of 6.7\arcdeg$\pm$2\arcdeg\ and decreases about linearly if the true distance is greater than the adopted distance of 700 pc.  This is in excellent agreement with the derivation from only the radial velocities of 6.5\arcdeg $\pm$ 2\arcdeg, the result presented in O'Dell et al.\@ (2007b). This is small enough to simplify the calculation of the distance.

\subsection{Complicating Factors in Relating Radial Velocity and Apparent Tangential Velocities}

There are two complicating factors that need to be considered carefully when deriving a kinematic distance to the nebula. The first is that the radial velocities derived from spectroscopy are most sensitive to the expansion of those regions of the nebula where the expansion direction is closest to the line of sight. In the case of the Ring Nebula, these are the polar regions of the nebula. The proper motions, on the other hand, will be more representative of the equatorial regions of the nebula, where the expansion direction is closest to the plane of the sky. In order to deal with this complication, it is necessary to adopt some model for the internal kinematics of the nebula. In this paper, we follow \citep{ode07b} in assuming a ``Hubble flow'' (all gas velocities strictly radial and linearly proportional to distance from the central star), with a nebular radius along the polar axis of 1.5 times that along the  major equatorial axis at \(\mathrm{PA} = 70\arcdeg\). Although this simple kinematic model seems to be broadly consistent with the observations, it is important to test it by searching for deviations from a pure Hubble flow, as emphasized by  Steffen et al.\@ (2008). 

The second complicating factor is that the proper motions measure a pattern speed, whereas the spectroscopy measures the physical speed of the gas. These are not necessarily the same thing, as elucidated in detail by Mellema (2004). In Appendix A, we derive the correction factor, \(\mathcal{R}\) (ratio of pattern speed to physical speed) as a function of two parameters: \(f_0\), which is the degree of concentration of the ionized gas at its outer edge, and \(\beta_*\), which is the rate of change of the stellar ionizing luminosity divided by the nebular expansion rate: \(\mathcal{R} = 2 f_0 /( 2 f_0 - 1 - \beta_*/3)\). The density profile found from fitting Cloudy models to nebular imaging and spectroscopy \citep{ode07b} shows a roughly linear increase with radius, implying \(f_0 \simeq 1.3\). The dynamic timescale \citep{ode07b} and the stellar evolution timescale \citep{blo95} are roughly equal at 4--5000 years, implying \(\beta_* \simeq -1\). Hence, we find \(\mathcal{R} \simeq 1.34\). 

\subsection{Derivation of a Distance to the Ring Nebula from its Radial and Tangential Velocities}

We can derive the distance to the Ring Nebula by assuming that the component of the spatial expansion  (\vs) in the plane of the sky is equal to the corrected tangential component, the former being determined from the radial velocity study \citep{ode07b} and the latter using the tangential angular motions determined in the present study. Equating these two components one has \vs\ x cos~$\theta$=4.74 $\mu$ x D x \(\mathcal{R}\)$^{-1}$, where units of arcsec/year, \kms, and parsecs are used. Within the assumption of homologous expansion, the distance from the central star drops out. Employing the velocity expansion relation (\vs =0.65x\phis\ \kms) determined from the radial velocities, the expansion rate of 0.23$\pm$0.10mas yr$^{-1}$ determined from the observed tangential angular velocity, and the value \(\mathcal{R}\)=1.34 from \S\ 5.3, one finds D=790$^{+500}_{-200}$ pc. This is larger, but within the uncertaintanties of the parallax determined value of 700$\pm ^{450}_{200}$ pc \citep{har07}. Since both methods determine the distance from the reciprocal of the measured quantity, it is illuminating to compare the measured parallax 1.42$\pm$0.55 mas with our predicted parallax 1.27$\pm$0.73 mas. Taking the average of these parallaxes (1.35$\pm$0.6), one determines a best recommended distance of 740$^{+500}_{-200}$ pc. This is only a small change from the parallax determined distance used previously, but, it is determined from two independent methods and is the value to now be preferred.

\subsection{The Knots in the Ring Nebula appear to be co-moving with the nebular gas}

The origin of the knots in the Ring, Helix, and other nearby PNe \citep{ode02} remains highly uncertain and an important diagnostic observation is whether the knots share the general expansion characteristics of the host PN or if they trail behind.  In a study of the radial velocities of the similarly nearly pole-on PN NGC 7293 (the Helix Nebula) \citep{mea98} argued that the knots were expanding, but at about one half the rate of expansion of the nebular material. This interpretation of the radial velocity data was disputed \citep{omm04} when the knot velocities were compared with several sources of nebular velocities, the conclusion being that the expansion rates of the  knots and the nebula are indistinguishable. That also appears to be the case for the Ring Nebula, since the plane of the sky expansion rate for the knots (0.24$\pm$0.11 mas yr$^{-1}$/\arcsec) is not distinguishably different from that for the nebular features (0.22$\pm$0.11 mas yr$^{-1}$/\arcsec).  This means that either the knots in the Ring Nebula move in a very different way than those in the Helix Nebula or that the Helix Nebula results based on radial velocities \citep{mea98} are incorrect. 

The images of the Ring Nebula made in the \htwo\ \twomic\ line (Speck et~al.\@ 2003) indicate that this emission arises from the same ionization bounded knots that produce the bright cusps seen in [N~II] and their dark central knots seen in [O~III].  The \htwo\ emission is probably excited by stellar radiation more energetic than 13.6 eV in the same manner as that in the Helix Nebula \citep{ode07a,hen07}. A lower spatial resolution Fabry Perot study \citep{hir04} at a resolution of 24 \kms\ also noted that the \twomic\ emission arose in these features. Hiriart (2004) used his data to derive a model in which the \twomic\  emission comes from a cylindrical shell seen almost along its rotational axis, which is fully consistent with the triaxial spheroid model ( O'Dell et~al.\@ 2007b) that has most of the material lying in an equatorial disk lying nearly in the plane of the sky.  Using the \htwo\ velocities and distances from Hiriart's study, one finds a ratio of expansion velocity to angular separation of 1.1 \kms/arcsec, which is within the determination errors of the value of 1.0 \kms/arcsec derived from the optical line study  \citep{ode07b} (c. f. \S\ 5.1), which had a velocity resolution of 7 \kms.  This agreement in radial velocity adds credence to the interpretation that the knots are moving at the same velocity as the nebular material.

On a small scale, many planetary nebulae show imhomogeneities in the form of dense molecular knots \citep{ode02}. In the case where the knot is well inside the global ionization front, there is
expected to be a locally divergent flow of ionized gas away from the
head of the knot, e.g., Bertoldi \&\ McKee (1990). This contradicts the assumption that all gas flows are radially away from
the central star, so the above analysis cannot be applied directly to
this case. However, the ionization front conditions on the knot should
be approximately D-critical (\(U - u = c_\mathrm{i}\), where \(
c_\mathrm{i}\) is the isothermal sound speed in the ionized gas).
Therefore,  one finds \(\mathcal{R} = 1 + \mathcal{M}^{-1}\), where
\(\mathcal{M} = u/c_\mathrm{i}\) is the Mach number of the ionized gas
at the surface of the knot on a radial line from the central star, as
measured in the rest frame of the star. In order to apply this
formula, it is necessary that both the proper motion observations
(which measure \(U\)) and the spectroscopic observations (which
measure \(u\)) are detecting the knots, rather than the diffuse
nebular emission. This requires a very high angular resolution, which
is not typically available from ground-based spectroscopy.

\subsection{How do the change of distance and new stellar photometry alter the picture of the state of evolution of the Central Star?}

In \S\ 2.2 we established that the visual brightness (V=15.65) of the central star was 0.1 magnitudes brighter than assumed previously \citep{ode07b}, but that the extinction was less (c$\rm _{H\beta}$=0.13, A$\rm _{V}$) than had been previously assumed (c$\rm _{H\beta}$=0.20, A$\rm _{V}$=0.43), so that this would not significantly change the previous result that the absolute luminosity of the central star is about 200 L$_{\sun}$, (these new numbers argue for its being 5\%\ fainter).  The largerer recommended distance (740 pc rather than 700 pc) argues that the star is an additional 11\%\ brighter. This means that at present there are no good grounds to question our previous conclusions about the evolutionary state of the central star, that is,  that it is a star of about 0.61-0.62 M$_{\sun}$ about 7000 years beyond the end of the Asymptotic Giant Branch wind.

\section{Summary and Conclusions}

We have obtained long time base images of NGC 6720, the Ring Nebula and from these have derived its expansion characteristics in the plane of the sky. Supplemental ground based spectra along the major and minor axis were also obtained. These data have allowed us to draw these conclusions.

1.  Scaled ratio images and the motion of individual features argues with considerable uncertainty for homologous expansion, that is, the rate of expansion varies linearly with the distance from the central star.

2. The dark knots seen against the bright nebular emission appear to be expanding at the same rate as the surrounding gas.

3. The spectra have allowed determination of the electron temperature in all of the ionization zones present in this high ionization PN and of the electron density in the lowest  (MIF) and medium (He$^{+}$) ionization zones.

4. Combining the tangential and radial velocity information, we have derived a new distance for the Ring Nebula of D=790$^{+450}_{-180}$ pc, slightly larger than the 700 $\pm ^{450}_{200}$ pc determined by parallax, although subject to revision because of correction for the difference in velocity of the MIF and the surrounding nebular gas. The best distance is now 740 $\pm ^{500}_{200}$ pc.

 \acknowledgments
 
 We are grateful to Yilen Gomez Maqueo Chew, who provided excellent assistance in obtaining the spectra at the San Pedro Martir Observatory. We thank Garrelt Mellema for useful comments on a draft of this paper. Part of CRO's support for work on this project came from grants GO 11231 and GO 11232 from the Space Telescope Science Institute. WJH acknowledges financial support from DGAPA-UNAM, Mexico, project IN110108.
 
{\it Facilities:} \facility{HST{(WFPC2)}}, {SPM}

\begin{appendix}

\section{Correction for difference between real and apparent motion}
\label{sec:relative-motion}

We derive here the relationship between the apparent propagation speed, \(U\), of an ionization front and the gas speed, \(u\), of ionized material immediately inside the front. This calculation is complementary to that of Mellema (2004), who employed a strictly local procedure, based on the ionization front jump conditions. Mellema's results are exact, but depend on a free parameter that must be supplied from the outside. We follow a global procedure, considering the nebula as a whole, allowing us to find a unique solution. On the other hand, we are forced to make some simplifying assumptions, which are nevertheless justified in most cases. 

We consider a spherically symmetric, ionization-bounded, pure-hydrogen nebula, surrounding a point source of ionizing radiation with an ionizing photon luminosity that varies with time as \(Q\H(t)\). The ionization front radius is \(R(t)\) and the total mass of ionized gas is \(M(t)\). The number density of hydrogen ions just inside the ionization front is \(n_0(t)\), whereas the mean density and mean square density within the ionized volume are given by \(\langle n \rangle\) and \(\langle n^2 \rangle\), respectively. We define a clumping factor, \(C = \langle n^2 \rangle / \langle n \rangle^2\) and a shell concentration factor, \(f_0 = n_0/\langle n \rangle\), which are both assumed to be constant with time. The proton mass is \(m\) and the Case~B recombination coefficient is \(\alpha\B\). We define the ionization front dynamic timescale as \(t\d = R/U\) and the stellar evolution timescale as \(t_* = Q\H / \dot{Q}\H\). \emph{Note that \(t_*\) is negative when the ionizing luminosity falls with time.}

Given these assumptions, the mass of ionized gas is given by 
\begin{equation}
  \label{eq:mass}
  M(t) = \frac{4\pi m}{3 f_0} n_0(t) R(t)^3
\end{equation}
and the dynamic ionization balance in the nebula is given by 
\begin{equation}
  \label{eq:ionbalance}
 Q\H(t) = \frac{4\pi \alpha\B C}{3 f_0^2}  n_0(t)^2 R(t)^3 + m^{-1}\frac{dM}{dt} ,
\end{equation}
where the first term on the right hand side represents the rate of recombinations in the nebula, while the second term represents the rate at which new particles are \textit{advected} through the ionization front. It is straightforward to show that the relative size of the second term is of order \(t_{\mathrm{rec}}/\min(t\d,|t_*|)\), where \(t_{\mathrm{rec}} = 1 / (\alpha\B n) \) is the recombination time. This ratio is rather small for typical nebular conditions, so we neglect the advective term in what follows, which is equivalent to assuming that the nebula passes through a sequence of static photoionization equilibrium states. With this approximation, we can take the time derivative of equation~(\ref{eq:ionbalance}), eliminating the density by use of equation~(\ref{eq:mass}), to obtain
\begin{equation}
  \label{eq:qdot}
  \frac{d \ln Q\H}{dt} = 2 \frac{d \ln M}{dt} - 3 \frac{d \ln R}{dt}, 
\end{equation}
which yields the rate of change of ionized mass as
\begin{equation}
  \label{eq:mdot}
  \frac{dM}{dt} = \frac{M}{2} \left(\frac{3}{t\d} + \frac{1}{t_*} \right). 
\end{equation}
On the other hand, we can also express the rate of change of ionized mass in terms of the relative velocity, \(U - u\), with which newly ionized gas leaves the ionization front: 
\begin{equation}
  \label{eq:massflow}
  \frac{dM}{dt} = 4\pi R^2 n_0 m (U - u) . 
\end{equation}
Equations~(\ref{eq:mdot}) and~(\ref{eq:massflow}) can be combined to find the correction factor, \(\mathcal{R} = U/u\), that relates the real and apparent expansion speeds:
\begin{equation}
  \label{eq:correct}
  \mathcal{R} = \frac{ 2 f_0 } { 2 f_0 - 1 - \beta_*/3 }, 
\end{equation}
where \(\beta_* = t\d/t_*\) is the ratio of the rate of evolution of the ionizing luminosity to the rate of nebular expansion. 

\subsection{Application to real Planetary Nebulae}

The ionizing luminosity of the central stars of planetary nebulae
first rises with time as the post-AGB atmosphere contracts towards
higher effective temperatures, and then falls swiftly when envelope
nuclear burning ceases, followed by a much slower decline as the white
dwarf interior cools \citep{blo95}. The overall
timescale for this sequence is set primarily by the core mass at the
end of the AGB phase (higher mass cores tend to evolve more rapidly),
but with a secondary dependence on the AGB mass-loss history. The
expansion timescale of the nebula generally increases monotonically
with the nebular age, so that the three phases in the luminosity
evolution (rise, swift decline, slow decline) produce three phases in
the nebular evolution: ionization (\(\beta_* > 0\)), recombination
(\(\beta_* < -3\)), re-ionization (\(-3 < \beta_* < 0\)). These phases
can be seen, for example, in the evolution of the mass of ionized gas
in the models of Natta \&\ Hollenbach (1998).

The detailed evolution of the nebula can be rather intricate, even
when only one-dimensional models are considered \citep{per04}, especially during the
high-luminosity phase when the fast stellar wind from the central star
is expected to be dynamically important in the inner regions of the
nebula. Nonetheless, so long as the nebula is optically thick to
ionizing radiation, the simple equation given above can be reliably
applied in order to correct expansion parallaxes measured from the
main nebula shell for cases where detailed tailored models, such as
those given in Sch\"onberner et~al.\@ (2005), are not
available.\footnote{Unfortunately, the models of
Sch\"onberner et~al.\@ (2005) do not cover the recombination and
re-ionization phases and therefore cannot be applied to evolved
nebulae such as the Ring and the Helix.} The [\ion{N}{2}] line is to
be preferred over [\ion{O}{3}] for both spectroscopic and proper
motion measurements, since it traces gas close to the ionization
front, giving the most reliable estimates of \(u\) and \(U\). Note
that the model given here does \textit{not} apply to density-bounded
nebulae, where the ionizing luminosity is high enough to ionize all
the circumstellar material. The lack of an ionization front means that
other features, such as shocks \citep{mel04}, must be used to
determine expansion parallaxes in such cases.

\subsection{Limiting cases}
\label{sec:limiting-cases}

It is instructive to consider some particular limits of the general equation~(\ref{eq:correct}). 

\paragraph{No stellar evolution}

If the ionizing luminosity, \(Q\H\), is constant, then \(\beta_* = 0\) and we have \(\mathcal{R} = 2 f_0 /(2 f_0 - 1) \). For a constant density nebula, \(f_0 = 1\) and we obtain \(\mathcal{R} = 2\), which is the well-known result for the evolution of a classical Str\"omgren sphere during the subsonic, D-type phase---the maximum ionized gas velocity is one half the expansion speed of the ionization front. On the other hand, for a thin, homogeneous ionized shell, of thickness \(h \ll R\), one finds \(f_0 \simeq R/(3 h)\) and \(\mathcal{R} \simeq 1 + \frac32 h/R\), so that the correction factor is much reduced, tending towards unity as the shell thickness tends to zero.

\paragraph{Rising ionizing luminosity}

If \(dQ\H/dt > 0\), then \(\beta_*\) is also positive, which will always lead to larger values of \(\mathcal{R}\) than in the constant luminosity case. The greatest increase is seen for a constant density nebula, whereas larger values of \(f_0\) tend to dampen the effect. As \(\beta_* \to 6f_0 - 3\), then equation~(\ref{eq:correct}) suggests \(\mathcal{R} \to \infty\). However, the equation is not valid in this limit, which corresponds to a highly supersonic weak-R type ionization front, since for such fronts the advective term would need to be retained in equation~(\ref{eq:ionbalance}). 

\paragraph{Falling ionizing luminosity}

If \(dQ\H/dt < 0\), then \(\beta_*\) is negative, which causes a reduction in the correction factor, \(\mathcal{R}\), with respect to the constant luminosity case. Irrespective of the value of the density factor \(f_0\), if \(\beta_* < -3\), then \(\mathcal{R} < 1\), so that the gas velocity is now greater than the expansion pattern speed. In this case, the nebula is bounded by a \emph{recombination front}, instead of an ionization front, and the mass of ionized gas falls with time.

\end{appendix}

\end{document}